\documentclass[twocolumn,showpacs,aps,prl,amssymb,amsfonts,amsmath,nofootinbib ]{revtex4}
\usepackage{amssymb}
\usepackage[vcentermath]{youngtab}
\usepackage{graphicx}
\newcommand{\Spec}{\mathop{\mathrm{Spec}}}
\newcommand{\Tr}{\mathop{\mathrm{Tr}}}
\newcommand{\bra}{\langle}
\newcommand{\ket}{\rangle}

\newcommand{\SU}{\mathop{\mathrm{SU}}}

\begin{document}

\title
{\bf The Pauli exclusion principle and beyond}
\author{Alexander A. Klyachko}

\affiliation{Faculty of Science, Bilkent University, Bilkent,
Ankara, 06800 Turkey}

\begin{abstract}
The Pauli exclusion principle can be stated as inequality
$\langle\psi|\rho|\psi\rangle\le 1$ for the electron density
matrix $\rho$. Nowadays it is replaced by skew symmetry of the
multi-electron wave function. The replacement leads to numerous
additional constraints on $\rho$, which are discussed in this
letter together  with their physical implications, in particular
for spin magnetic moment of a multi-electron system.
\end{abstract}


\maketitle

Recall, that the original Pauli exclusion principle claims that no
quantum state can be occupied by more than one electron
\cite{Pauli}. This sets a bound on expected number of electrons in
any given state $\psi$
\begin{equation}\langle\psi|\rho|\psi\rangle\le 1.\label{Pauli}
\end{equation} Here $\rho=\sum_{ij}\bra\Psi|a_i^\dag
a_j|\Psi\ket|\psi_i\ket\bra\psi_j|$
is Dirac's density matrix
of $N$-electron system in state $\Psi$.
Soon after Pauli discovery, Heisenberg and Dirac \cite{Sudarshan}
replaced it by skew symmetry of the multi-electron wave function
\begin{equation}\Psi\in\wedge^N
\mathcal{H}_r\subset \mathcal{H}_r^{\otimes N},\label{skew_sym}
\end{equation}
where the wedge stands for skew symmetric tensors built  on the
one-electron spin-orbital space $\mathcal{H}_r$ of dimension $r$.

In this Letter I first briefly review a recent solution of a
longstanding problem about impact of the Dirac-Heisenberg
replacement on the electron density matrix
\cite{Klyachko06,Klyachko08}. It goes far beyond the original
Pauli principle and leads to numerous {\it extended Pauli
constraints\/} independent of (\ref{Pauli}). In the rest of the
paper I discuss some of their physical implications.

The problem has a long and complicated  history. A difference
between the two versions of Pauli principle is best seen in
Heisenberg's {\it exchange term\/} in electron energy. By adding
this term Fock drastically improved Hartree mean field theory
which initially complies the exclusion principle only in its
original form. The exchange energy plays also crucial r\^{o}le in
formation of covalent bonds, as well as in many other two-electron
effects. This urged L\"owdin and Shull \cite{Lowdin} thoroughly
analyze two-electron wave functions, whose structure is closely
related to eigenvectors and eigenvalues of the corresponding
density matrix $\rho$. L\"owdin \cite{Lowdin55} coined for them
the terms {\it natural orbitals\/} and {\it natural occupation
numbers\/}. It turns out that the natural orbitals of a
two-electron system split into pairs $\psi_i^{(1)},\psi_i^{(2)}$
and the total state takes the form $\Psi=\sum_i
a_i\psi_i^{(1)}\wedge\psi_i^{(2)}$. The orbitals
$\psi_i^{(1)},\psi_i^{(2)}$ have the same occupation numbers
$\lambda_i^{(1)}=\lambda_i^{(2)}=|a_i|^2$, and therefore spectrum
of $\rho$ is doubly degenerated.

Beyond two electrons $\wedge^2\mathcal{H}_r$ and two holes
$\wedge^{r-2}\mathcal{H}_r$ \cite{Ruskai70}
the progress was very slow. In early seventies
Borland and Dennis \cite{B-D}
tried out numerically the simplest such system
$\wedge^3\mathcal{H}_6$
and discovered the following constraints on
the natural occupation numbers $\lambda=\Spec\rho$
arranged in decreasing order
$\lambda_1\ge\lambda_2\ge\cdots\ge\lambda_r$ and normalized to the
number of electrons $N=\Tr\rho$:
\begin{equation}\lambda_1+\lambda_6=\lambda_2+\lambda_5 =\lambda_3+\lambda_4=1
,\quad \lambda_4\le\lambda_5+\lambda_6.\label{B-D}
\end{equation}
A rigorous proof has been found about the same time by
M.B.~Ruskai, who published it only recently \cite{Ruskai}.

These (in)equalities may challenge our physical intuition. They
allow exactly {\it one\/} electron in {\it two\/} symmetrical
natural orbitals and clearly supersede the original Pauli principle,
which in this setting reads $\lambda_1\le 1$. This example revealed
a dramatic effect of the skew symmetry condition on structure of the
density matrix and provoked the authors' emotional comment
\begin{quote}
{``We have no apology for consideration of such a special case.
The general $N$-repre\-sentability problem is so difficult and yet
so fundamental for many branches  of science that each concrete
result is useful in shedding light on the nature of general
solution."}
\end{quote}
Shortly afterward Peltzer and Brandstatter \cite{Peltzer} published
a {\it false\/} solution, claiming  that two-electrons, two-holes,
and the Borland--Dennis system $\wedge^3\mathcal{H}_6$ are
exceptional, and in all other cases the Pauli constraint
$\lambda_1\le 1$ is the only one. After that the problem has stalled
for more than three decades.\footnote{Mostly because of lack a
proper theoretical insight, that emerged much latter
\cite{Klyachko98,  Klyachko02, B-Sj00}, and inconclusive results of
numerical studies.}

Now, with a complete solution at hand, it may be appropriate to
return back to physics and look for possible manifestations of the
extended Pauli constraints.

Let me first give a sample of results for a three-electron system
$\wedge^3\mathcal{H}_r$ of even or infinite rank $r$:
\begin{eqnarray}
\label{sym_orb} &\lambda_{k+1}+\lambda_{r-k}\le 1,\quad 0\le k<r;&\\
\label{quadrpl} &\begin{array}{ll}
 \lambda_2+\lambda_3+\lambda_4+\lambda_5\le 2,&\quad
\lambda_1+\lambda_3+\lambda_4+\lambda_6\le2,\\
\lambda_1+\lambda_2+\lambda_5+\lambda_6\le 2,&
\quad\lambda_1+\lambda_2+\lambda_4+\lambda_7\le2;
\end{array}\quad&\\
\label{inf_ext}
&\lambda_1+\lambda_2+\lambda_4+\lambda_7+\lambda_{11}+\lambda_{16}+\cdots\le
2.&
\end{eqnarray}
The inequalities  (\ref{sym_orb}) improve the original Pauli
condition $\lambda\le 1$. Due to normalization $\Tr\rho=3$ they turn
into Borland--Dennis equations (\ref{B-D}) for $r=6$. The quadruple
of inequalities (\ref{quadrpl}) provides a full set of constraints
for $r\le7$. A proper infinite extension of the last inequality in
the quadruple is given by (\ref{inf_ext}) where the differences
between successive indices are natural numbers $1,2,3,4,\ldots$.
Similar extensions for the other three constraints can be obtained
by adding to them a tail of the series (\ref{inf_ext}) starting with
the term $\lambda_{11}$.

A complete set of constraints heavily depends on the rank $r$ and
the number of particles. Currently they are known for all systems
of rank $\le10$. To give an idea of complexity of the problem note
that for systems $\wedge^3\mathcal{H}_{10}$,
$\wedge^4\mathcal{H}_{10}$, $\wedge^5\mathcal{H}_{10}$ the
constraints amount to $93$, $125$, $161$ independent inequalities
respectively.

\vspace{2mm} {\it Pinned state effect.\/} Recall, that the Pauli
principle is a purely {\it kinematic\/} constraint on available
states of a multi-electron system. Not even a minuscule violation
has been detected so far, in spite of numerous and increasingly
sophisticated attempts \cite{VIP06}. Therefore whenever a dynamical
trend is in conflict with Pauli constraints, the latter would
prevail and the system eventually will be trapped in the boundary of
the manifold of allowed states. This can manifest itself in
degeneration some of the extended Pauli inequalities into equations.
In this case the system and its state vector will be called {\it
pinned\/} to the degenerate Pauli inequalities. Since a pinned
system is primary driven by Pauli kinematics, rather than by
Hamiltonian dynamics, it should be robust and remain pinned under a
reasonably small variation of the Hamiltonian. This stability gives
us a chance to detect pinned states  and in a sense makes them real.
A pinned system is essentially a new physical entity with its own
dynamics and kinematics.

The concept can be illustrated by a string pendulum. For a small
initial horizontal impulse at the lowest point its mass is confined
to a circle. However, if the energy is big enough to raise the mass
above the suspension point but is not sufficient to neutralize
gravitation at the apex, then at some moment the string constraint
becomes irrelevant and the mass switches into a free
trajectory.

As a more relevant example, consider the first excited state of
beryllium atom of spin $S=1$ and $S_z=1$. Here are its natural
occupation numbers calculated from 10 spin-orbitals in the lowest
three shells $1s$, $2s$, $2p$ \cite{Nakata01}\vspace{-3mm}

{\footnotesize
\begin{equation}\label{Be}
\begin{array}{lllll}
1.000 000,& 0.999 995,& 0.999 287,& 0.999 284, & 0.000 711,\\
0.000707,& 0.000 009, &0.000 007, &0.000 000, &0.000 000.
\end{array}
\end{equation}}
Observe that the first orbital is completely filled and the last
two are empty, i.e.
the state is pinned to the initial Pauli constraints $0\le
\lambda\le 1$. This kind of degeneration is very common.
In its ultimate form it has been elevated  to the {\it Aufbau\/}
principle and incorporated into Hund's rules.

The completely filled and empty orbitals are inactive, and
therefore we effectively deal with a reduced system
$\wedge^3\mathcal{H}_7$, in which all orbitals are partially
filled. The extended Pauli constraints for this system amount to
the quadruple of inequalities (\ref{quadrpl}). Note that the
exclusion principle itself $\lambda_1\le 1$ doesn't enter into the
list explicitly, but can be derived by taking sum of the second
and third inequalities that gives
$\lambda_1+\lambda_6-\lambda_7\le 1$. One can infer from this that
the Pauli degeneration $\lambda_1=1$ for a system of this format
would pin it  to the last  three constraints (\ref{quadrpl}), that
still fall short of  $\lambda_1=1$. Actual calculation for
beryllium data (\ref{Be}) shows that one of these three
inequalities turns into exact equation
\begin{equation}\label{Be_pin}
\lambda_1+\lambda_2+\lambda_4+\lambda_7=2,
\end{equation}
and the other two are exact within a rounding error.

This equality stretches the Pauli principle to its limit, and
therefore restricts the system's kinematics.
Recasting  it into the form
\begin{equation}\label{Be_cnstr}
(a_1^\dag a_1+a_2^\dag a_2+a_4^\dag a_4+a_7^\dag a_7)\Psi=2\Psi
\end{equation}
we get a {\it selection rule\/} for Slater determinants
$[i,j,k]=\psi_i\wedge\psi_j\wedge\psi_k$
that enter into the decomposition $\Psi=\sum c_{ijk}[i,j,k]$. Such
a determinant should include  two natural orbitals from the set
$\{1,2,4,7\}$, and  one from its complement $\{3,5,6\}$. As a
result the pinned system splits into two components
$\wedge^2\mathcal{H}_4\otimes\mathcal{H}_3$, spanned by the above
quadruple and triplet, and containing respectively  two and one
electron. The components can be readily identified with two
spin-up electrons, and one spin-down.

The selection rules coming from the other two pinned inequalities
bound the decomposition to four Slater determinants
\begin{equation}\label{Be_stat}\Psi=\alpha[1,2,3]+\beta[1,4,5]+\gamma[1,6,7]+\delta[2,4,6]
\end{equation}
with the amplitudes, up to phase factors, set by  equations
$$
\begin{array}{rrr}
|\alpha|^2=\lambda_3,&|\beta|^2=\lambda_5, &|\gamma|^2=\lambda_7,
\\ |\alpha|^2+|\delta|^2=\lambda_2,&
|\beta|^2+|\delta|^2=\lambda_4,& |\gamma|^2+|\delta|^2=\lambda_6.
\end{array}
$$
This gives us an insight into  structure of the pinned beryllium
state whose form (\ref{Be_stat}) should be stable under a
perturbation. It corresponds to the type VII in Schouten
classification of 3-vectors $\Psi\in\wedge^3\mathcal{H}_7$
\cite{Schouten31}.

A similar consideration can be applied to the Borland-Dennis system
$\wedge^3\mathcal{H}_6$ in which every state is pinned by two {\it
equations\/} (\ref{B-D}). They reduce its internal kinematics to
that of three qubits $\mathcal{H}_2\otimes
\mathcal{H}_2\otimes\mathcal{H}_2$, each spanned by a couple of
symmetric orbitals $\psi_{i},\psi_{j}$, $i+j=7$. In this
interpretation the respective pairs of occupation numbers
$\lambda_i,\lambda_{j}$ become spectra of three reduced states of
$\Psi\in\mathcal{H}_2\otimes \mathcal{H}_2\otimes\mathcal{H}_2$,
and the inequality (\ref{B-D}) turns into  Higuchi-Sudbery-Szulc
compatibility condition \cite{Higuchi03}.

\vspace{2mm} {\it Spin-orbit interaction versus Pauli
constraints.\/} In the above setting we were unable to address
specific spin effects buried in join spin-orbital space
$\mathcal{H}_r=\mathcal{H}_\ell\otimes\mathcal{H}_s$. Actually the
orbital $\mathcal{H}_\ell$ and spin $\mathcal{H}_s$ degrees of
freedom play very different r\^{o}le. The former, via Coulomb
interaction,  are primary responsible for dynamics, whereas  the
latter, disregarding  a small relativistic correction, are purely
kinematic. In this approximation the total spin is a good quantum
number, that single out a specific component in Weyl's
decomposition
\begin{equation}\label{Weyl}
\wedge^N(\mathcal{H}_\ell\otimes\mathcal{H}_s)=\sum_{|\nu|=N}
\mathcal{H}_\ell^\nu\otimes\mathcal{H}_s^{\nu^t}.
\end{equation}
Here $\mathcal{H}_\ell^\nu$ is an irreducible representation of
the orbital group $\SU(\mathcal{H}_\ell)$ defined by Young diagram
$\nu$ and $\mathcal{H}_s^{\nu^t}$ is the representation of spin
group $\SU(\mathcal{H}_s)$ with transpose Young diagram $\nu^t$.
The electron spin group is $\SU(2)$ and in this case $\nu$ is a
two-column Young diagram. For example, in the ground closed-shell
state of beryllium atom $\nu=\text{\tiny $\yng(2,2)$}$\,, while
for the excited state of spin one $\nu=\text{\tiny
$\yng(2,1,1)$}$. In general the total spin $S$ is equal to
half-difference of lengths the two columns of $\nu$.

For spin resolved state
$\Psi\in\mathcal{H}_\ell^\nu\otimes\mathcal{H}_s^{\nu^t}$ the
Pauli constraints amount to inequalities between orbital
$\lambda_i$ and spin $\mu_j$ natural occupation numbers normalized
to traces $N$ and $1$ respectively \cite{Klyachko08}.
As a toy example consider three electrons in $d$-shell
($\dim\mathcal{H}_\ell=5$) in low spin configuration
$\nu=\text{\tiny $\yng(2,1)$}$\,
where the  constraints are as follows
\cite[\textit{6.1}]{Klyachko08}
\begin{eqnarray*}
&\lambda_1+\frac12(\lambda_4+\lambda_5)\le 2,\\
&
\begin{array}{ll}
\mu\le 3-2(\lambda_1-\lambda_2),\quad&\mu\le
3-2(\lambda_2-\lambda_3),\\
\mu\ge 2(\lambda_1-\lambda_3)-3,\quad&\mu\ge
4\lambda_1-2\lambda_2+2\lambda_4-7.
\end{array}&
\end{eqnarray*}
Here $\mu=\mu_1-\mu_2$ is spin magnetic moment in Bohr magnetons
$\mu_B$ and I set $g=2$ for the electron gyromagnetic
factor.

The first inequality in the list improves the Pauli condition
$\lambda_1\le 2$ for the number of electrons in an orbital.

The second line provides a kinematic upper bound on the spin
magnetic moment. This contrasts sharply with prevailing  opinion
that the reduction of magnetic moment  is a  dynamical effect
caused by spin-orbit coupling.\footnote{See for example
\cite[p.98]{Koebler06}: ``If spin-orbit interaction is zero free
spin magnetism is realized. This is meant by saying that the
orbital moment is quenched."}

One has to be careful with the  lower bound for $\mu=\mu_1-\mu_2$
given in the last line. Recall, that $\mu_1,\mu_2$ are
probabilities to find the system in spin-up and spin-down states
with respect to some  natural direction determined by leading
eigenvector of the spin density matrix. It may not coincide with
the direction of magnetic field $H$ used for measurement of the
moment. However, for negligible spin-orbit coupling the natural
direction is free to orient itself along the field to produce  the
maximal moment $\mu=\mu_1-\mu_2$. Otherwise the coupling might
reduce this ideal value
to the level that breaks the Pauli constraint. This would be a
signature of an appreciable spin-orbit interaction.

This example provides a key to an old puzzle about reduction of
magnetic moment of a transition metal atom in a crystal, relative to
its value in free space. I will focus below on body-centered cubic
form of iron ($\alpha$-Fe), stable at normal condition, using a
recent high precision $\gamma$-ray diffraction study by Jauch and
Reehuis \cite{Jauch07}.

The study support an earlier finding \cite{Pipkorn64}, bases on
measurement of a M\"ossbauer shift,  that the electron configuration
of $3d$-shell in iron crystal is $d^7$ rather than
$d^6$ as for free atom. Such a modification of metallic $3d$-shell
is pretty common, albeit not universal.  It reduces the maximal spin
magnetic moment per atom from $4\,\mu_B$ to $3\,\mu_B$, while the
experimental saturation moment is $2.22(1)\,\mu_B$.

Recall, that in a  crystal field of cubic symmetry the orbital
moment is quenched and $d$-shell splits into two irreducible
components of dimension 3 and 2 called $t_{2g}$ and $e_{g}$
subshells.
The orbital density matrix retains the crystal symmetry, and
therefore reduces to scalars $n_t$ and $n_e$ on the above subshels.
As a result the orbital occupation numbers
$\lambda=(n_t,n_t,n_t,n_e,n_e)$, $3n_t+2n_e=7$ depend only on one
parameter $n_t$, where I expect $n_t\ge n_e$.

\begin{figure}[h]
\includegraphics[width=7cm]{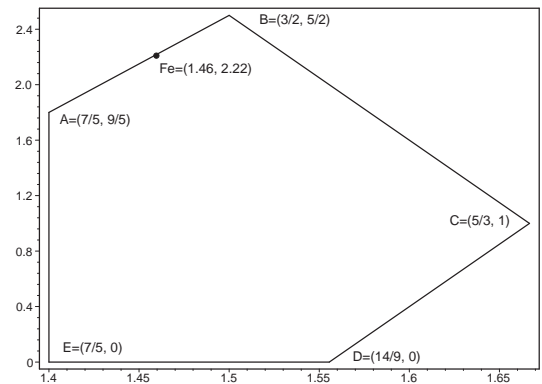}
\caption{\footnotesize \label{fig1} Pauli constraints on spin
magnetic moment ($\mu_B$) for 7 electrons in $d$-shell in cubic
crystal field versus the occupation number $n_t$ of a $t_{2g}$
orbital. All points within the pentagon $ABCDE$ are admissible. A
black dot represents experimental data for iron. } \vspace{-0.5cm}
\end{figure}

The Pauli constraints on spin magnetic moment $\mu$ versus
the orbital occupation number $n_t$ for high spin $d^7$
configuration are shown in Fig.~\ref{fig1}. The plot also includes
the experimental data for iron $\mu=2.22$, $n_t=1.458(7)$. The
latter value is deduced from the relative number of $d$-electrons
$62.5(3)\%$ in $t_{2g}$-subshell \cite{Jauch07}.

The data suggest that the observed magnetic moment is the maximal
one  kinematically allowed by the electron density, i.e. the iron
$d$-shell is pinned to the boundary segment $[A,B]$ given by
equation $\mu=7n_t-8$. Let me stress again that the constraint on
spin magnetic moment $\mu\le7n_t-8$ comes from the
kinematics of $d$-shell and has nothing to
do with spin-orbit coupling. One can interpret it as an ultimate
result of a multi-electron ``exchange interaction", though the term
suggests a dynamical origin of the effect, while in fact it is
purely kinematic.

The constraints shown in Fig.~\ref{fig1} are actually derived from a
system of $55$ inequalities between orbital $\lambda_i$ and spin
$\mu_j$ natural occupation numbers. The pentagon $ABCDE$
is a projection of a multidimensional polytope that
includes all spin occupation numbers $\mu_i$, rather than just the
total magnetic moment $\mu=3\mu_1+\mu_2-\mu_3-3\mu_4$. It turns out,
however, that the boundary segment $[A,B]$ has unique pull back
$[\widetilde{A},\widetilde{B}]$ in the bigger polytope

\vspace{2mm} {\footnotesize
$$\widetilde{A}=\left[\frac75,\frac75,\frac75,\frac75,\frac75\left|
\frac35,\frac15,\frac15,0\right]\right.,\quad \widetilde{B}=
\left[\frac32,\frac32,\frac32,\frac54,\frac54\left|
\frac34,\frac14,0,0\right]\right.$$} where the first 5 components
represent the orbital occupation numbers and the rest are spin
ones. This allows to recover spin occupation numbers $\mu_i$ of
iron $d$-shell
\begin{equation}\label{spin_occc_nmb}(0.69,\;0.23,\;0.08,\; 0).
\end{equation}
The small probability $0.08$  to have negative magnetic moment
$-1\,\mu_B$ can be seen in a spin density plot
\cite[p.170]{Neut_Scatt06} built from polarized neutron scattering
data. I wonder if the data contain enough  information to recover
the spin occupation numbers
and verify the values (\ref{spin_occc_nmb}).

\vspace{2mm} {\it Magnetovolume effect.\/} Pressure is an effective
tool to change electronic structure, though apparently it remains
detached  from spin. This makes the observed reduction of magnetic
moment under pressure a mystery, known as magnetovolume effect
\cite{Iota07}. I use it below to test robustness of pinned iron
states.

For $\alpha$-iron the magnetic moment is governed
by the pinning equation $\mu=7 n_t-8$,
at least when the pressure is not too high. If at some point the
constraint would cease to exists, then the frozen degrees of
freedom will be released and change the dynamics of the system. No
such change is observed until a transition from bcc to hcp crystal
structure has started. The magnetic moment at this point is about
$1.9\,\mu_B$ \cite{Iota07} and the electron density should be
close to spherical one $n_t\approx n_e\approx 1.4$ to accommodate
the two incompatible symmetries. The transition actually passes
through a phase coexistence interval. A linear extrapolation to
the point where bcc phase vanishes gives the  magnetic moment
about $1.8\,\mu_B$. This is consistent with the evolution along
the pinning edge $[B,A]$.

The effect is more tricky in a face-centered cubic form of iron
stable at high temperature
or low density.
It has two competing
spin states, first conjectured by Weiss \cite{Weiss63}. The one
with high moment
$\mu\approx2.5\,\mu_B$ at some critical pressure collapses into a
low spin one $\mu\approx1\,\mu_B$.

Recall that $2.5\,\mu_B$ is the maximal magnetic moment
kinematically allowed for cubic iron, provided $n_t\ge n_e$.
Assuming the latter condition we infer that the high spin state
should be pinned to the point $B$ and remains trapped at this
point until one of the two pinning equations $\mu=7n_t-8$ or
$\mu=16-9n_t$ is released. As the above scenario suggests, at some
critical pressure the first constraint is eventually freed and the
state evolves along the segment $[B,C]$ into the low spin state
$C$, where it will be stabilized until the second pinning
constraint $\mu=16-9n_t$ is relaxed. Then it slides along the path
$[C,D]$ into a nonmagnetic phase. This picture qualitatively fits
the calculated \cite{Mgn_Inst_The} and the observed
\cite{Mgn_Inst_Exp} evolutions and gives a hint about possible
origin of the two spin states.

May be I have to conclude with a disclaimer that this letter is
not about iron magnetism, used here as a testing ground for new
concepts and methods originated from a recent advance in
understanding of quantum kinematic constraints.
Until now the  kinematic effects were
largely neglected or treated as dynamical ones. The confusion was
clearly recognized by chemists \cite{Cox95}, and the above examples
may support their objection.
In the end, the role of Hamiltonian in some cases may reduce to
pinning a state to a certain kinematic constraints which
qualitatively shape the system's behavior.

\end{document}